\documentclass{siamltex}

\usepackage{amsfonts,amssymb,amsmath,graphicx,color}

\def\eps{\varepsilon}

\def\RR{\mathbb{R}}  
\def\EE{\mathbb{E}}

\def\<{\langle} \def\>{\rangle}

\title{arclength parametrized Hamilton's equations\\ for the
  calculation of instantons}

\author{T. Grafke \footnotemark[2]
\and R. Grauer \footnotemark[3]
\and T. Sch{\"a}fer \footnotemark[4]
\and E. Vanden-Eijnden \footnotemark[5]}

\begin{document}

\maketitle

\renewcommand{\thefootnote}{\fnsymbol{footnote}}
\footnotetext[2] {Department of Physics of Complex Systems, Weizmann Institute of Science, Rehovot 76100 Israel.}
\footnotetext[3] {Theoretische Physik I, Ruhr-Universit{\"a}t Bochum, Universit{\"a}tsstr. 150, D-44780 Bochum, Germany.}
\footnotetext[4] {Department of Mathematics, College of Staten Island 1S-215, 2800 Victory Blvd., Staten Island, New York 10314.}
\footnotetext[5] {Courant Institute, New York University, 251 Mercer Street, New York, New York.}

\renewcommand{\thefootnote}{\arabic{footnote}}

\begin{abstract}
  A method is presented to compute minimizers (instantons) of action
  functionals using arclength parametrization of Hamilton's
  equations. This method can be interpreted as a local variant of the
  geometric minimum action method (gMAM) introduced to compute
  minimizers of the Freidlin-Wentzell action functional that arises in
  the context of large deviation theory for stochastic differential
  equations. The method is particularly well-suited to calculate
  expectations dominated by noise-induced excursions from
  deterministically stable fixpoints. Its simplicity and
  computational efficiency are illustrated here using several
  examples: a finite-dimensional stochastic dynamical system (an
  Ornstein-Uhlenbeck model) and two models based on stochastic partial
  differential equations: the $\phi^4$-model and the stochastically
  driven Burgers equation.
\end{abstract}

\section{Introduction}
\label{sec:intro}

Finding the minimizer of action functionals is a fundamental problem
that appears in a variety of areas in mathematics and science. For
example, it arises in the context of large deviation theory (LDT),
where probabilities and expectations over the solutions of stochastic
differential equations (SDEs) can be estimated in the small noise
limit via minimization of the Freidlin-Wentzell action
\cite{freidlin-wentzell:1998}. This has led to the development of
numerical tools specifically designed to tackle this task: the string
method (for gradient fields)
\cite{e-ren-vanden-eijnden:2002,e2007simplified}, the minimum action
method \cite{e-ren-etal:2004,fogedby-ren:2009}, the adaptive minimum
action method (aMAM) \cite{zhou2008adaptive} and the geometric minimum
action method (gMAM)
\cite{heymann-vanden-eijnden:2008,heymann2008pathways,heymann-vanden-eijnden:2008b}
are a few examples of techniques to calculate the minimizer of the
Freidlin-Wentzell action, also called instanton. One difficulty one is
faced in these calculations is that the speed of the instanton can be
very inhomogeneous: for example it can vanish at critical locations
making its total duration infinite. In these situations, it is
computationally advantageous to replace the time-parametrization of
the instanton by some other, geometrically motivated parametrizations,
such as e.g.  arclength parametrization. This strategy is a the core
of the string method as well as aMAM and gMAM. The aim of the present
paper is to revisit these methods and propose alternative ways to
calculate instantons by focusing on the Hamiltonian counterpart of the
Euler-Lagrange equation of gMAM.

To be specific, we will consider a system of SDEs in $\mathbb{R}^n$
with a drift vector~$b(x)$ and a diffusion matrix $a = \sigma\sigma^T$
weighted by a small parameter $\epsilon$ characterizing the strength
of the noise:
\begin{equation}
  \label{eq:sde}
  dX^\epsilon(t) = b(X^\epsilon(t))dt + \sqrt{\epsilon}\sigma dW(t).
\end{equation}
For simplicity, we will assume that the diffusion matrix $a$ is
constant (but not necessarily diagonal) -- the generalization of the
ideas below to non-constant $a$ is straightforward.  From the theory
of large deviations~\cite{varadhan2008large,freidlin-wentzell:1998} it
is known that in the limit as~$\epsilon\to 0$, the solutions
to~\eqref{eq:sde} that contribute most to the probability of an event
or the value of an expectation are likely to be close to the minimizer
of the Freidlin-Wentzell action functional $S_T$ subject to
appropriate boundary conditions. This action functional is given by
\begin{equation}
\label{eq:Freidlin-Wentzell_action}
S_T(x) = \int_0^T\,L(x,\dot x)\,dt
\end{equation}
with the Lagrangian
\begin{equation}
  L(x,\dot x) = \frac{1}{2}\langle \dot x-b(x),a^{-1}(\dot
  x-b(x))\rangle\,
  \equiv \frac{1}{2}|\dot x -b(x)|_a^2.
\end{equation}
Here $\<x,y\>$ denotes the Euclidean scalar product between the
vectors $x$ and $y$ and we introduced the norm induced by $a$, $|x|_a
= \sqrt{\langle x,a^{-1}x\rangle}$ -- if $a$ is the identity, this is
simply the Euclidean length.

The instanton, i.e. the path that minimizes the action
(\ref{eq:Freidlin-Wentzell_action}), can be found by solving the
corresponding Euler-Lagrange equations
\begin{equation}
\label{eq:Euler-Lagrange_original}
\frac{d}{dt}\frac{\partial L}{\partial \dot x}=\frac{\partial L}{\partial x}
\end{equation}
with the appropriate boundary conditions. Thus, in the context of
large deviation theory
the estimation of probabilities or expectations can be reduced to the
solution of the deterministic
system~\eqref{eq:Euler-Lagrange_original}. The question then becomes
how to do this efficiently, for example in situations where one is
searching for solutions to~\eqref{eq:Euler-Lagrange_original} with one
end-point fixed at a stable critical point of $\dot x = b(x)$ -- this
case is relevant e.g. for the estimation of expectations with respect
to the stationary distribution of~\eqref{eq:sde}. This is the main
topic of this paper, which is organized as follows. In
section~\ref{sec:hamilton} we derive the arclength parametrized
Hamilton's equations that are equivalent
to~\eqref{eq:Euler-Lagrange_original}. We also show how these
equations can be used to estimate the stationary probability
distribution of the SDE~\eqref{eq:sde} via calculation of the
quasipotential from LDT (section~\ref{sec:interp}), as well as
expectations with respect to this distribution
(section~\ref{sec:expect}). Finally we give a few special solutions of
these equations (section~\ref{sec:special}). In section~\ref{sec:algo}
we propose a simple and efficient algorithm for the numerical
integration of the arclength parametrized Hamilton's equations. This
algorithm is then tested on several illustrative examples: a finite
dimensional Ornstein-Uhlenbeck process that is amenable to analytical
solution (section~\ref{sec:OU}), the $\phi^4$-model
(section~\ref{sec:phi4}) and the randomly forced Burgers equation
(section~\ref{sec:burgers}). Some concluding remarks are given in
section~\ref{sec:conclusion}.

\section{Hamilton's equations with arclength parametrization}
\label{sec:hamilton}

Instead of solving (\ref{eq:Euler-Lagrange_original}), we can also
compute the instanton in the corresponding Hamiltonian framework. The
Hamiltonian~$H$ is the Legendre transform of the Lagrangian~$L$, 
\begin{equation}
H = \langle \dot x,  p \rangle - L, \qquad p = \frac{\partial L}{\partial \dot x}.
\end{equation}
and the Hamil\-ton's equations for the instanton are
\begin{equation}
  \label{eq:hamilton}
   \dot x = \frac{\partial{H}}{\partial p}, \qquad 
   \dot p = - \frac{\partial{H}}{\partial x} \,.
\end{equation}
As mentioned before, the main idea of gMAM is that, in a wide range of
situations, the instanton can be computed much more efficiently by
using a parametrization that is different than the time~$t$. Let us
consider a reparametrization of the form
\begin{equation}
   \label{eq:reparametrization}
    t = t(s), \qquad \bar x(s) = x(t(s)), \qquad \bar p(s) = p(t(s))
\end{equation}
and assume that $t'(s)>0$. By chain rule, we have then
\begin{displaymath}
\bar x ' = t' \frac{\partial{H}}{\partial \bar p}, \qquad 
\bar p '= - t' \frac{\partial{H}}{\partial \bar x} , \qquad t' = t'(s)
\end{displaymath}
and we can write
\begin{equation}
  \label{eq:hamilton_gmam_general}
  \lambda \bar x' = \frac{\partial{H}}{\partial \bar p}, \qquad 
  \lambda \bar p' = - \frac{\partial{H}}{\partial \bar x}, \qquad 
  \lambda = \frac{1}{t'(s)}.
\end{equation}
One possibility to choose the parametrization is to require that
$\|\bar x'\| = C$, where $\|\cdot\|$ is a suitable norm and $C$ is a
constant giving the length of $x$ in this norm. This fixes the value
of the function $\lambda$ in~\eqref{eq:hamilton_gmam_general}, which
can be thought of as a Lagrange multiplier introduced to enforce the
constraint~$\|\bar x'\| = C$. Once $\lambda$ is known, $t(s)$ and its
inverse $s(t)$ can then be obtained from the last equation
in~\eqref{eq:hamilton_gmam_general}. It is often natural to use the
norm induced by the correlation matrix $a$, hence to define
$\|x'\|\equiv |x'|_a$. For simplicity, we call this {\em arclength
  parametrization} even if $a$ is not the identity. Note that, in
principle, norm (and parametrization) can be chosen freely and,
indeed, other choices are possible. In the present work, however, we
concentrate on arclength parametrization in contrast to the original
time parametrization. Details of the implication of other
parametrizations (and norms) will be studied elsewhere.

For diffusion-driven systems, the Hamiltonian is given by
\begin{equation}
  \label{eq:hamiltonian_diffusion}
  H = \frac{1}{2}\langle p, ap \rangle + \langle b, p\rangle,
\end{equation}
and \eqref{eq:hamilton} read explicitly
\begin{equation}
\label{eq:hamilton_time_parametrization}
\dot x = a p+b, \qquad \dot p = -(\nabla b)^T p
\end{equation}
where $\nabla b$ is the tensor with entries $(\nabla b)_{i,j} =
\partial b_i/\partial x_j$.  Correspondingly the first two equations
in \eqref{eq:hamilton_gmam_general} become
\begin{equation}
\label{eq:hamilton_gmam_diffusion}
\lambda \bar x' = a\bar p+b, \qquad \lambda \bar p' = - (\nabla
b)^T\bar p,
\end{equation}
In this case, the parametrization by
arclength $|\bar x'|_a=C$ directly leads to $\lambda = |a\bar
p+b|_a/C$ and we can
write~\eqref{eq:hamilton_gmam_diffusion} as
\begin{equation}
  \label{eq:hamilton_gmam_diffusion_1}
  \bar x' = \frac{C}{|a\bar p+b|_a}(a\bar p+b), \qquad 
  \bar p' = -\frac{C}{|a\bar p+b|_a} (\nabla b)^T\bar p.
\end{equation}
The norm appearing in the denominator can be expressed as
\begin{equation}
  \label{eq:consH}
  \begin{aligned}
    |a\bar p+b|_a^2 &= \langle a\bar p+b,a^{-1}(a\bar p+b)\rangle =
    \langle a\bar p+b, \bar p+a^{-1}b\rangle \\
    &= \langle a\bar p,\bar p\rangle + 2\langle \bar p,b \rangle +
    |b|_a^2 = 2H + |b|_a^2 = 2E + |b|_a^2,
  \end{aligned}
\end{equation}
where we used the property that since the Hamiltonian $H$ does not
depend explicitly on time, energy is conserved and the points on the
graph $(\bar x,\bar p)$ need to satisfy the constraint $H=E$ everywhere. This
leads to yet another equivalent form
of~\eqref{eq:hamilton_gmam_diffusion_1}:
\begin{equation}
  \label{eq:hamilton_gmam_diffusion_1b}
  \bar x' = \frac{C}{\sqrt{2E + |b|_a^2}}(a\bar p+b), \qquad 
  \bar p' = -\frac{C}{\sqrt{2E + |b|_a^2}} (\nabla b)^T\bar p.
\end{equation}
These equations are valid in particular in the case when $H=E=0$,
which, as we will see in sections~\ref{sec:interp}
and~\ref{sec:expect} is the relevant one to compute probabilities and
expectations with respect to the stationary distribution
of~\eqref{eq:sde}. In the sequel, we will focus on this case, that is,
we will study
\begin{equation}
  \label{eq:gmam_c}
  x' = \frac{C}{|b|_a}(ap+b), \qquad p' = -\frac{C}{|b|_a} (\nabla b)^Tp.
\end{equation}
where we dropped the bar to simplify notations - we will stick to this
convention in the sequel.  Note that these equations seem singular at
the critical points where $b(x)=0$. This is not the case, however, as
$p=0$ at these points, too. But it will complicate the numerical
integration of (\ref{eq:gmam_c}) since, typically, minimizing paths
start from and may go through critical points - this issue will be
dealt with in section~\ref{sec:algo}. Also note that the constant
$C=|x'|_a$ giving the length of the path is, usually, not known
\textit{a~priori}. Finally, note that if $a$ were not constant, an
additional term $-\tfrac12(C/|b|_a) \sum_{i,j} \partial a_{i,j} /\partial x_k p_i
p_j $ should be added to the equation for $k$-component of $p$.

\subsection{Derivation of~\eqref{eq:gmam_c} from the geometric action
  and link with the quasipotential}
\label{sec:interp}

A key quantity in Freidlin-Wentzell theory is the quasipotential,
defined as
\begin{equation}
  \label{eq:quasipotential}
  V(x_1,x_2) = \inf_{T>0} \inf_{x}  S_T(x)
\end{equation}
where $x_1$ and $x_2$ are two arbitrary points in $\RR^n$ and the
infimum over $x$ is taken over all the paths that satisfy $x(0)=x_1$,
$x(T) = x_2$. The quasipotential permits to estimate various long-time
properties of~\eqref{eq:sde} in the limit as $\eps\to0$. For example,
if $x_0$ is the unique stable fixpoint of $\dot x = b(x)$, and
under suitable conditions such that the solutions to~\eqref{eq:sde} are
ergodic with respect to the stationary probability density function
$\rho_\eps(x)$, then
\begin{equation}
  \label{eq:2}
  \rho_\eps(x) \asymp e^{-\eps^{-1} V(x_0,x)}
\end{equation}
where $\asymp$ indicates that the ratio of the logarithms of both sides
tends to 1 as $\eps\to0$. In \cite{heymann-vanden-eijnden:2008}, it
was shown that the quasipotential~\eqref{eq:quasipotential} can be
represented equivalently as
\begin{equation}
  \label{eq:quasipotentialb}
  V(x_1,x_2) = \inf_{x}  \hat S(x)
\end{equation}
where the infimum is taken over the paths that satisfy $x(0)=x_1$,
$x(1) = x_2$ and we introduced the geometric action
\begin{equation}
  \label{eq:gmam_action}
  \hat S(x) = \int_0^1 \hat L(x,x') ds, \qquad \hat L(x,x') = |x'|_a |b(x)|_a - \<
  x',b(x)\>_a.
\end{equation}
This quantity bears his name from the property that it is left
invariant by any reparametrization of the path.

We show now by explicit calculation that arclength parametrized
Hamilton's equations \eqref{eq:gmam_c} are equivalent to the
Euler-Lagrange equations associated with (\ref{eq:gmam_action}). These
Euler-Lagrange equations are given by
\begin{equation}
  \label{eq:euler_lagrange_1}
  \left(\frac{\partial \hat L}{\partial x'}\right)' = \frac{\partial
    \hat L}{\partial x},
\end{equation}
which we can write in a more explicit form as
\begin{equation}
  \label{eq:euler_lagrange_2}
  \left(\frac{a^{-1} x'}{|x'|_a}|b(x)|_a - a^{-1} b(x)\right)' 
  = \frac{|x'|_a}{|b(x)|_a}(\nabla
  b (x))^T a^{-1} b(x) - (\nabla b(x))^T a^{-1} x'.
\end{equation}
Let
\begin{equation}
  \label{eq:definition_of_p}
  a p = \frac{x'}{|x'|_a}|b(x)|_a - b(x).
\end{equation}
Then (\ref{eq:euler_lagrange_1}) can be written as
\begin{equation}
  \label{eq:equation_for_p}
  p' = \frac{|x'|_a}{|b(x)|_a}(\nabla b (x))^T a^{-1} b(x) - (\nabla
  b(x))^T a^{-1} x'.
\end{equation}
If we solve (\ref{eq:definition_of_p}) in $x'$, and use this
expression in (\ref{eq:equation_for_p}), the resulting system of
equations is exactly~\eqref{eq:gmam_c}.

\subsection{Use of~\eqref{eq:gmam_c} in the
  computation of expectations}
\label{sec:expect}

Assume again that $\dot x = b(x)$ has a single stable fixpoint at
$x=x_0$ and suppose that we want to estimate the following expectation
with respect to the stationary density of the process.
\begin{equation}
  \label{eq:13}
  I = \int_{\RR^n} \exp\left(-\epsilon^{-1} f(x)\right) \rho_\eps(x) dx
\end{equation}
where $f(x)$ is some function. Then as $\eps\to0$
\begin{equation}
  \label{eq:14}
  I \asymp \exp\left ( -\eps^{-1} \inf [S(x) + f(x(1))]\right),
\end{equation}
where the infimum is taken over all paths such that $x(0)=x_0$. It is
easy to see that the minimizer of this variational problem
solves~\eqref{eq:gmam_c} with the boundary
conditions
\begin{equation}
  \label{eq:boundary_x_p}
  x(0) = x_0, \qquad p(1) = - \nabla f(x(1)).
\end{equation}
Note that these boundary conditions differ from the boundary
conditions that are commonly used in applications of the original
gMAM. In those applications, {\em two points}, e.g. two stable
fixpoints $x_0$ and $x_1$ are prescribed. In contrast, when computing
expectations as shown above, one of these boundary conditions is
replaced by $p(1) = - \nabla f(x(1))$.  It is possible to adapt gMAM
to handle such boundary conditions, for example by making the end
point $x(1)$ variable. However, as shown in section~\ref{sec:algo},
working with~\eqref{eq:gmam_c} offers a much simpler way to handle
boundary conditions of the type~\eqref{eq:boundary_x_p}.

\subsection{Special solutions of~\eqref{eq:gmam_c}}
\label{sec:special}

If $p=0$, then (\ref{eq:gmam_c}) reduces to
\begin{equation}
  \label{eq:xprime_parallel_b}
  x' = \frac{C}{|b(x)|_a} b(x),
\end{equation}
meaning that $x'\parallel b(x)$ and point in the same direction. This
solution corresponds to a reparametrized version of a deterministic
trajectory, which is the solution of $\dot x = b(x)$. Another $p$ that
solves $H=0$ with $H$ given by~\eqref{eq:hamiltonian_diffusion} is $p
= -2a^{-1} b(x)$, for which the equation for $x$ (\ref{eq:gmam_c})
reduces to
\begin{equation}
  \label{eq:xprime_antiparallel_b}
  x' = -\frac{C}{|b(x)|_a} b(x),
\end{equation}
meaning that $x'\parallel b(x)$ and points in the opposite
direction. This special solution, however, is not always consistent
with the second equation in (\ref{eq:gmam_c}). Indeed, differentiating
$p = -2a^{-1} b(x)$ gives
\begin{equation}
  \label{eq:equation_pprime}
  p' = -2a^{-1} \nabla b(x) x' =  2 \frac{C}{|b(x)|} a^{-1} \nabla b(x) b(x),
\end{equation}
where we used (\ref{eq:xprime_antiparallel_b}) to get the second
equality. This equation is consistent with the second equation in
(\ref{eq:gmam_c}) \textit{iff} $\nabla b = (\nabla b)^T$,
i.e. \textit{iff} the drift term is gradient, $b(x) = - \nabla U(x)$
for some potential $U(x)$. Thus, solutions of (\ref{eq:gmam_c}) that
follow deterministic paths in reverse can only be observed in gradient
systems.

Another class of systems for which we can find special solutions
of~\eqref{eq:gmam_c} that are not~\eqref{eq:xprime_parallel_b} are
those for which~\cite{freidlin-wentzell:1998}
\begin{equation}
  \label{eq:specialb}
  b(x) = -a \nabla U(x) + d(x)
\end{equation}
where $d(x)$ is such that $\nabla \cdot d = 0$ and $d \cdot \nabla U =
0$. It can easily be checked that in this case the solutions to
\eqref{eq:sde} are ergodic with
respect to the following stationary
probability density
\begin{equation}
  \label{eq:3}
  \rho_\epsilon(x) = C^{-1}_\epsilon e^{-2U(x)/\epsilon}
\end{equation}
where $C_\eps = \int_{\RR^n} e^{-2U(x)/\epsilon} dx$ is a
normalization factor. In that case a special class of solutions
to~\eqref{eq:gmam_c} are those such that $p = -2\nabla U$ and $x$ satisfies
\begin{equation}
  \label{eq:xprime_antiparallel_b2}
  x' = \frac{C}{|b(x)|_a} (a\nabla U + d), 
\end{equation}
as can be checked by direct substitution.

\section{Algorithmic aspects}
\label{sec:algo}

Using the arclength parametrized Hamilton's equations
(\ref{eq:gmam_c}) gives us a direct and numerically efficient
iterative method to find the instanton that fits the boundary
conditions (\ref{eq:boundary_x_p}): Assume that we are given the $k$-th
iteration $(x^{(k)},p^{(k)})$ together with an approximation $C^{(k)}$
of the path length as approximation to the solution of
(\ref{eq:gmam_c}) with the boundary conditions
(\ref{eq:boundary_x_p}). To compute the next iteration
$(x^{(k+1)},p^{(k+1)})$ and $C^{(k+1)}$ we proceed in the following
way:
\begin{enumerate} 
\item Use the known approximate solution $x^{(k)}=x^{(k)}(s)$ and
  $C^{(k)}$, together with the known final condition $p(1)= -\nabla
  f(x^{(k)}(1))$, to solve the equation for $p$ backwards in the
  arclength-parameter $s$ to obtain the next iteration $p^{(k+1)}$.
\item Use this computed approximate solution $p^{(k+1)}=p^{(k+1)}(s)$
  and $C^{(k)}$, together with the initial condition $x(0)=x_0$, in
  the equation for $x$ to obtain the next iteration $x^{(k+1)}$.
\item Compute the length of $x^{(k+1)}$ to obtain the next iteration
  of~$C^{(k+1)}$.
\end{enumerate}
If this iteration scheme converges to a fixpoint for $x$ and $p$, then
the constraint $|x'|_a=C$ will be automatically satisfied by this
fixpoint.  It may not be satisfied during the iterations, however, in
which case in step 3 it may be useful to also reparametrize
$(x^{(k+1)},p^{(k+1)})$ to enforce the constraint
$|x'|_a=\text{const}$. At convergence, the solution will also satisfy
$H=0$, since the first equation in~\eqref{eq:gmam_c} implies that
$|b|_a^2 = |ap+b|^2_a$ and hence $0=|p|^2_a + 2\<b,p\> \equiv 2H$. A
similar method (with a parametrization using an exponentially scaled
time) was suggested by Chernykh and Stepanov
\cite{chernykh-stepanov:2001} for the computation of instantons in
stochastic Burgers equation. We will discuss this example in more
detail in section \ref{sec:applications}.

It is worth stressing that the implementation of the method is
extremely simple: The solution of (\ref{eq:gmam_c}) requires only
forward (or backward) propagation of the initial (or final) value,
hence existing codes can be easily adapted to be used in this new
method. In contrast, gMAM usually requires the design of entirely new
code. The locality in memory of the method presented here is
particularly well-suited for parallelization: For large-scale
problems, it was successfully implemented in a massively parallel
setup, using graphics processing units (GPUs) to speed up the
computation (this will be presented elsewhere). In contrast, global
minimization techniques such as Newton or Quasi-Newton methods are
more difficult to implement in such a setting. These aspects
additionally simplify the scaling to a large number of degrees of
freedom. The largest computations for stochastic partial differential
equations presented here, with $2048$ grid-points in space and $16384$
in time take less than $10$ minutes in Matlab (on an Intel Xeon
E5-1620 with 3.60GHz) to successfully converge to the minimizer.


\section{Applications and Examples} 
\label{sec:applications}

Let us now present several examples and applications of the new
method. In these examples, we focus on the exit from a stable fixed
point and apply the iterative method described section~\ref{sec:algo}
in two ways: First, using Hamilton's equations in the original
time-parametrization given by (\ref{eq:hamilton_time_parametrization}):
\begin{displaymath}
\dot x = ap+b, \qquad \dot p = -(\nabla b)^Tp
\end{displaymath}
and second using the arclength parametrized Hamilton's
equations~\eqref{eq:gmam_c}:
\begin{displaymath}
x' = \frac{C}{|b|_a}(ap+b), \qquad p' = -\frac{C}{|b|_a} (\nabla b)^Tp.
\end{displaymath}
In all cases, we consider the escape from a stable fixpoint $x_0$ that
can be used as initial condition for $x$ in the above equations and we
designate a final condition for~$p$ either by fixing it or through the
second boundary condition in~\eqref{eq:boundary_x_p} for some
specific~$f$. Note that, in the time-parametrized case, since $x_0$ is
a stable fixpoint, it takes infinite time to escape it. This makes the
iteration numerically difficult as, in practice, one needs to choose a
$t_{\text{min}}$ and then decrease it towards $ -\infty$ until
convergence of the computed minimum of the action.  This problem does
not occur when one works with arclength-parametrization, which is an
immediate numerical advantage of this method.

\subsection{Application to a linear Ornstein-Uhlenbeck model}
\label{sec:OU}

As a first example, consider a linear Ornstein-Uhlenbeck model for
which the drift $b$ is of the form
\begin{equation}
  b(x) = -Bx,
\end{equation}
with $x\in\mathbb{R}^2$ and the matrix $B$ not necessarily normal. In
this case, a lot of analytical information about the minimizer can be
obtained in the following
way~\cite{freidlin-wentzell:1998,gardiner:2009}: Let $C$ be the
solution of
\begin{equation}
  BC + CB^T =  a
\end{equation}
with $a$ being the diffusion tensor from above. Then $\epsilon C$ is
the covariance of the equilibrium distribution for the SDE, which is
zero-mean Gaussian and exists as long as $x=0$ is a stable fixpoint
of $\dot x = - Bx$. The path of maximum likelihood of escape from $0$
is the time-reversed of the solution to
\begin{equation}
  \label{eq:trv}
  \dot x = - C B^T C^{-1} x.
\end{equation}
This result is consistent with the fact that the time-reversed form of
the SDE~\eqref{eq:sde} with $b(x) = -B x$ is
\begin{equation*}
  dX^\eps(t) = - C B^T C^{-1} X^\eps(t) dt + \sqrt{\epsilon} \sigma dW.
\end{equation*}
and it corresponds to picking
\begin{equation}
  \label{eq:H0ex}
  a p = Bx +C B^T C^{-1} x
\end{equation}
as solution to $H=0$. It can be checked that, if we
insert~\eqref{eq:H0ex} in the first equation
in~\eqref{eq:hamilton_time_parametrization}, this equation reduces to
\eqref{eq:trv}, and that the second equation
in~\eqref{eq:hamilton_time_parametrization} is also satisfied.  Since
we can compute the minimizer analytically, this model is a good start
to compare the two methods (Hamilton's equations with
time-parametrization vs Hamilton's equations with
arclength-parametrization) and to study convergence. As a concrete
example, we chose
\begin{equation*}
  B = 
  \begin{pmatrix}
    -1 & 4\\
    -4 & -1
  \end{pmatrix}
\end{equation*}
and $x(t=-\infty)=0$, $p(t=0)=(4,1)$ as a test case. Note that, in
this case, $B$ is normal and, therefore, the analytical solution is
even simpler: The minimizing path is the time-reversed of the
trajectory that spirals into the fixpoint at the origin and is given
by
\begin{align*}
  p(t) &= \exp \left( -B^T t \right) p_0\\
  x(t) &= -\left(B+B^T\right)^{-1}\! p(t)\,.
\end{align*}

For this system we implemented the algorithm described in
section~\ref{sec:algo} to solve both the equations of motion of the
original Hamiltonian \eqref{eq:hamilton_time_parametrization} and of
the geometric parametrization \eqref{eq:gmam_c}. Propagation of the
initial conditions is realized with a simple first order explicit
Euler scheme.

\begin{figure}[htb]
  \begin{center}
    \includegraphics[width=0.75\textwidth]{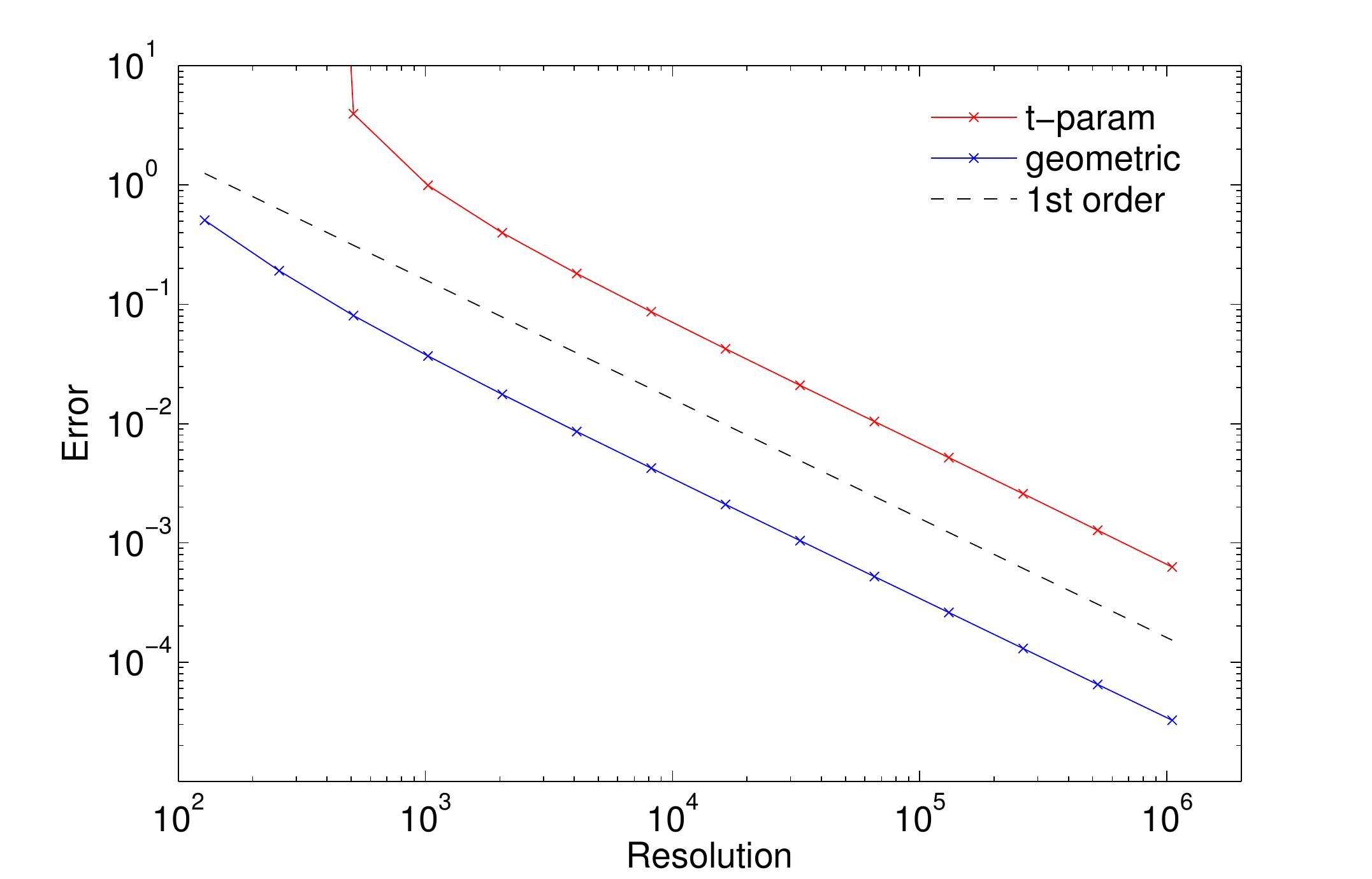}
  \end{center}
  \caption{Convergence of $x(t=0)$ to the analytical solution for the
    first test problem. Due to the choice of the time integration
    scheme, both methods converge to the analytical solution in first
    order. For a given resolution, the reparametrized method is more
    accurate by a factor $>10$.}
  \label{fig:convergence}
\end{figure}

As shown in figure~\ref{fig:convergence}, with identical resolution
the proposed method is closer to the analytical solution by a factor
$> 10$ (for a given accuracy, the proposed method needs roughly a
factor 10 grid-points less). Due to the choice of the integration
scheme, both methods converge with first order. The difference to the
analytical solution is measured by comparing the endpoints of the
paths, $x(t=0)$ (due to different parametrizations, differences
between other points than the endpoint are harder to define and
calculate).

\begin{figure}[htb]
  \begin{center}
    \includegraphics[width=0.75\textwidth]{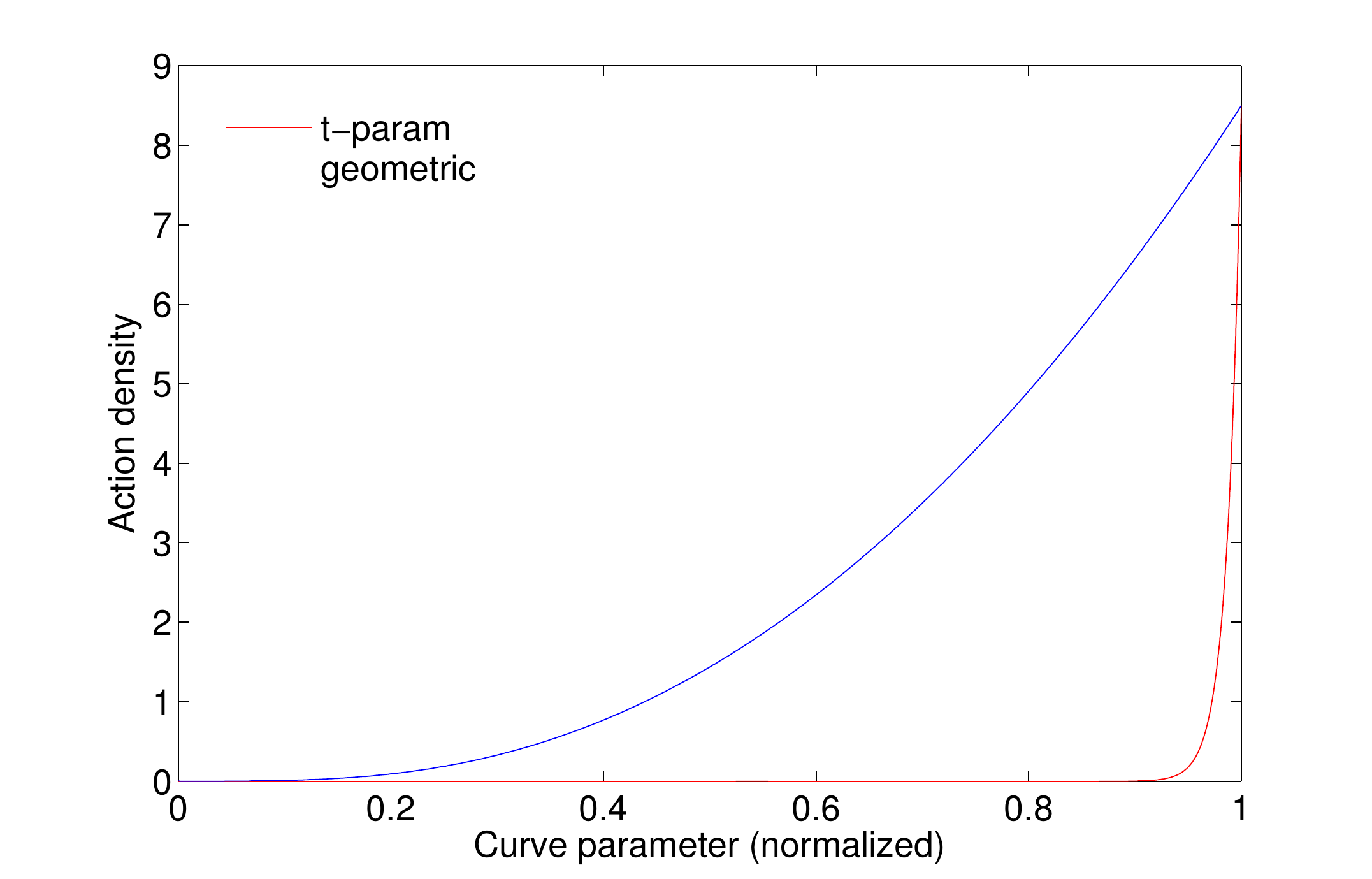}
  \end{center}
  \caption{Action density against curve parameter. For the simple
    iterative method, the curve parameter corresponds to the
    (normalized) physical time $t$. In the reparametrized method,
    regions with large action density are resolved better. }
  \label{fig:action}
\end{figure}

The action density over the curve parameter is concentrated on times
close to $t=0$ for the time parametrization, while it gets stretched
out to cover a significant part of the temporal domain in the geometric case
(see figure~\ref{fig:action}). This behavior is desired, because it
offers more grid points in the regions with critical dynamics.


\subsection{Application to the $\phi^4$ model} 
\label{sec:phi4} 

Consider the following SPDE
\begin{equation}
  \label{eq:20}
    \phi_t = \phi_{xx} - \phi^3 + \sqrt{2\epsilon}\, \eta(x,t),
    \qquad x \in (-L,L)
\end{equation}
with Dirichlet boundary conditions $\phi(-L) = \phi(L) = 0$, and where
$\eta(x,t)$ is a spatio-temporal white-noise. It can be
shown~\cite{faris1982large} that this equation is well-posed, and that
its equilibrium distribution is formally given by
\begin{equation}
  \label{eq:distrib}
  \mu (\phi) \propto \exp\left(-\epsilon^{-1} \int_{-L}^L (\tfrac12 |\phi_x|^2 +
    \tfrac14 |\phi|^4) dx\right).
\end{equation}
The right way to interpret this measure is via its Radon-Nikodym
derivative with respect to the measure of the Brownian bridge (i.e. by
using the measure of the Brownian bridge as reference and writing the
density of the equilibrium distribution of~\eqref{eq:20} with respect
to this measure -- see e.g. \cite{reznikoff2005invariant}):
\begin{equation}
  \label{eq:21}
  \frac{d\mu}{d \nu} = \exp\left(-\tfrac14\epsilon^{-1} \int_{-L}^L
    |B|^4) dx\right),
\end{equation}
where $\nu$ denotes the measure of the Brownian bridge $B(x)$ on
$[-L,L]$, i.e. of the Gaussian process with mean zero and covariance
\begin{equation}  \label{eq:22}
  \EE B(x) B(y) = \min(x+L,y+L) - \frac{(x+L)(y+L)}{2L}.
\end{equation}
In the large deviation regime (i.e. in the limit when $\eps\to0$), we
can ask what is the most likely configuration on the invariant measure
such that $\phi(0) = \phi_0$. This is the solution of
\begin{equation}
  \label{eq:23}
  \frac{d^2\phi}{dx^2} = \phi^3, \quad \phi(0) = \phi_0, \quad \phi(\pm
  L) = 0 
\end{equation}
In the limit as $L\to\infty$, this solution is simply
\begin{equation}
  \label{eq:24}
  \phi(x) = \frac{\phi_0}{1+\phi_0 |x|/\sqrt{2}}.
\end{equation}
We can also compute
\begin{equation}
  \label{eq:25}
  \EE \exp\left(\epsilon^{-1} \lambda \phi(0)\right),
\end{equation}
where $\lambda>0$ is a parameter and the expectation is taken on the
invariant measure. The configuration on the invariant measure that
contributes the most to this expectation is the solution
to~\eqref{eq:23} with the boundary condition $\phi_x(0+)= -\phi_x(0-)
= -\lambda$ and $\phi(\pm L) = 0$:
\begin{equation}
  \label{eq:24b}
  \phi(x) = \frac{\sqrt{\lambda}}{1+\sqrt{\lambda} |x|/\sqrt{2}}.
\end{equation}
The path of maximum likelihood to get dynamically to~\eqref{eq:24}
or~\eqref{eq:24b} is then the time reversed of the solution to
\eqref{eq:20} with $\eps=0$ (deterministic flow), and \eqref{eq:24}
or~\eqref{eq:24b} as initial condition. 

A similar story holds if we consider \eqref{eq:20} with periodic
boundary conditions, which are easier to implement numerically. In
this case the solution to~\eqref{eq:23} with the boundary condition
$\phi_x(0+) = -\phi_x(0-) = -\lambda$ and $\phi(L) = \phi(-L)$ gives
the relevant configuration for the periodic case.

In this example, we implemented the iterative scheme, both for the
equations of motion in the time-parametrized and the geometric variant
by using a second order Heun integration in the curve parameter. The
$x$-dimension is resolved with $N_x=512$ grid-points, using Fourier
transforms to calculate the spatial derivatives.

\begin{figure}[htb]
  \begin{center}
    \includegraphics[width=0.75\textwidth]{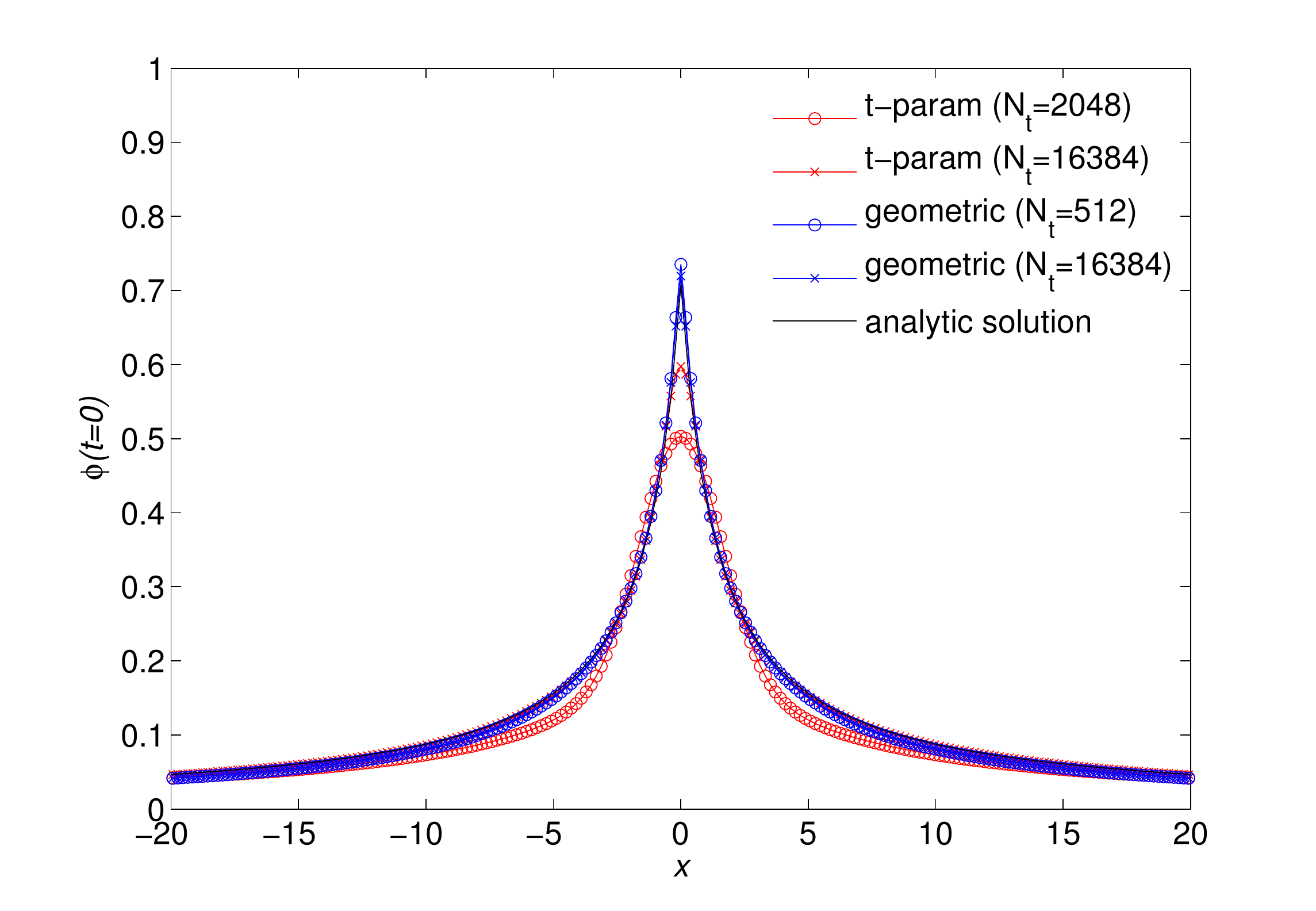}
  \end{center}
  \caption{Zoom to the origin for the final configuration in the
    proposed method and the time-parametrized method, in comparison to
    the analytical solution for the model system \eqref{eq:20}. The
    geometric solution at a moderate resolution of $N_t=512$ approximates the
    analytical solution better than the time-parametrized solution at
    a higher resolution of $N_t=16384$.}
  \label{fig:phi4}
\end{figure}

Figure~\ref{fig:phi4} compares, for different resolutions, the
numerical solutions of the final configuration of the minimizer to the
model~\eqref{eq:20} to the analytic solution for the periodic case. At
a resolution of $N_t=512$, the geometric method already outperforms
the time-parametrized variant with a considerably higher resolution of
$N_t=16384$ in terms of accuracy.

\subsection{Application to the stochastically driven Burgers
  equation}
\label{sec:burgers}

Consider the stochastically driven Burger's equation with periodic
boundary conditions:
\begin{equation}
  \label{eq:burgers}
   u_t = B(u) + \xi(x,t), \qquad B(u) = -uu_x + \nu u_{xx}.
\end{equation}
In this example, we consider a noise term $\xi$ that has a finite
correlation length in space, but is still $\delta$-correlated in
time. Hence we can write
\begin{equation}
 \label{eq:burgers_corr}
 \langle \xi(x,t) \xi(x',t') \rangle = \chi(x-x')\delta(t-t')
\end{equation}
and for the correlation function in space, we assume the following
form in Fourier space
\begin{equation}
 \label{eq:burgers_space_corr}
\hat\chi(\omega) = \omega^2\,{\mathrm{e}}^{-\omega^2/2} {\mathcal{H}}(\omega_c-|\omega|),
\end{equation}
where ${\mathcal{H}}$ denotes the Heaviside-function. This form
implies that the stochastic driving occurs only at frequencies up to a
cut-off frequency $\omega_c$. In this way, (\ref{eq:burgers}) is a
simple model for turbulence \cite{cardy-falkovich-etal:2008}: The
system is stochastically driven at larger scales by the noise term
$\xi$. This energy is transported via the non-linearity $u u_x$ from
large scales to small scales where energy is dissipated due to the
presence of the diffusive term $\nu u_{xx}$.

We consider the velocity field starting at rest for $t\rightarrow
-\infty$ and we are interested in noise realizations that lead to
strong negative gradients at $t=0$. This setup is similar to the
analytic treatment in \cite{balkovsky-falkovich-etal:1997}. We choose
as our observable
\begin{equation}
 \label{eq:burgers_observable}
  {\mathcal{O}}(u) = \int w(x) u_x(x)\, dx, 
  \qquad w(x) = \frac{1}{\sqrt{2\pi \sigma^2}}{\mathrm{e}}^{-x^2/(2\sigma^2)}.
\end{equation}
For this case, our Hamiltonian equations for the reparametrized fields
$u(x,s)$, $p(x,s)$ can be written as
\begin{equation}
  \label{eq:burgers_H_equations}
  \begin{cases}
    \displaystyle u_s =\frac{\|u_s\|_{\chi}}{\|B(u)\|_{\chi}}\left(\int
      \chi(x-x') p(x')dx'
      +B(u)\right) \\[8pt]
    \displaystyle p_s=  -
    \frac{\|u_s\|_{\chi}}{\|B(u)\|_{\chi}} \left(up_x+\nu
      p_{xx}\right).
  \end{cases}
\end{equation}
The initial condition for the velocity field $u$ is $u(s=0) = 0$ for
the final condition for the auxiliary field $p$ is $p(s=1) = -w_x$ as
can be seen from (\ref{eq:burgers_observable}) using integration by
parts. The norm $\|\cdot\|_{\chi}$ is given by the inverse of $\chi$
on its support (which is compact in Fourier domain), hence
\begin{equation}
\label{eq:norm_correlated}
\|f\|_{\chi} = \left(\left\langle f, {\mathcal{F}}^{-1}\left(\hat\chi^{-1}\hat f\right)\right\rangle\right)^{1/2},
\end{equation}
where, ${\mathcal{F}}^{-1}$ denotes the inverse Fourier transform
operator and $\hat f=\mathcal{F} f$ is the Fourier transform of $f$.

For comparison with previous work \cite{chernykh-stepanov:2001,
  grafke-grauer-schaefer:2013}, it is useful to contrast
\eqref{eq:burgers_H_equations} with the corresponding instanton
equations in the original parametrization (using the time $t$ as
parameter)
\begin{equation}
  \label{eq:burgers_time}
  \begin{cases}
    \displaystyle u_t + uu_x -\nu u_{xx} = \int \chi(x-x')p(x')dx' \\
    \displaystyle  p_t +up_x +\nu p_{xx} = 0.
  \end{cases}
\end{equation}
Note that the same equations can also be obtained using the
Martin-Siggia-Rose/Jan\-ssen/de Dominicis formalism
\cite{martin-siggia-rose:1973,janssen:1976,dedominicis:1976,phythian:1977}.
A main problem in solving~\eqref{eq:burgers_time} with the boundary
conditions $u=0$ at $t\rightarrow-\infty$ and $p(0-,x) = -w_x$
consists in the fact that the fixpoint of the system is only reached
in the limit $t\rightarrow -\infty$ and not in finite time. When
Chernykh and Stepanov computed a numerical solution to the system
above, they used a combination of two clever tricks to mitigate this
difficulty: First, they used self-similar properties of the
heat-kernel to design a coordinate transform that leads to an
exponential rescaling in time, second they used the linearization of
the system around zero in order to replace the boundary condition
$u=0$ by $u=\chi_0 p$ (where the constant $\chi_0$ can be computed
from the correlation function $\chi$). In the following we show that
the geometric reformulation of the instanton equations leads to a
natural rescaling of time that allows for direct efficient numerical
solution, which is furthermore transferable to similar situations
without any modification.

We implemented the iterative algorithm outlined in section
\ref{sec:algo}, employing a second order Heun integration of the
geometric equations of motion \eqref{eq:burgers_H_equations} and
compare the results to a similar iterative algorithm, but integrating
the time-parametrized equations of motion \eqref{eq:burgers_time},
instead. The space variable is resolved with $N_x=512$ grid-points,
and all derivatives in space are calculated via Fourier
transforms. For the calculations presented here we took $\sigma=0.1$
and $\nu = 1/2$ in~\eqref{eq:burgers_observable}

\begin{figure}[tb]
  \begin{center}
    \includegraphics[width=0.7\textwidth]{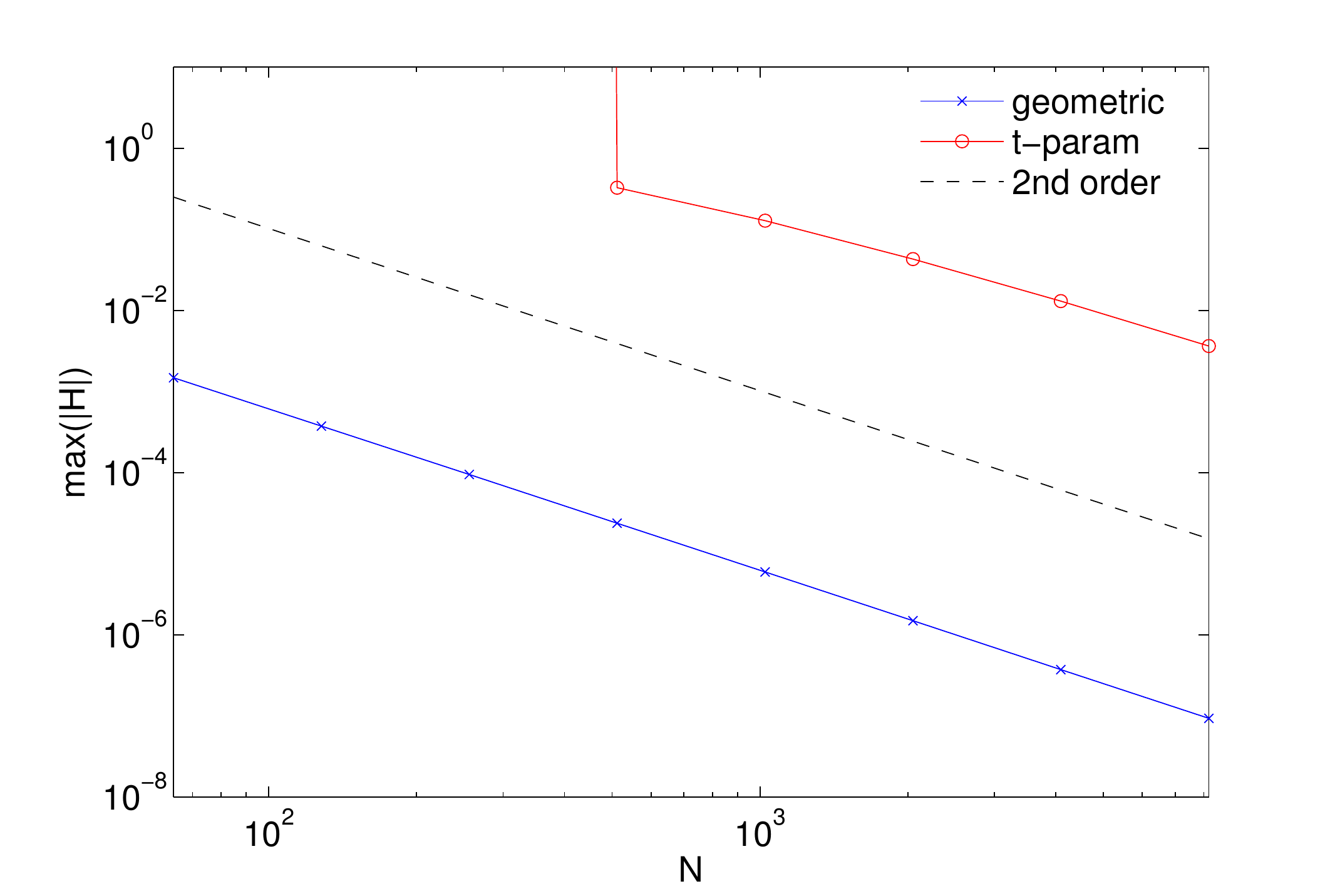}
  \end{center}
  \caption{Convergence of the maximum of the Hamiltonian along the
    minimizer. Both methods converge second order. For the smallest
    tested resolution of $N_t=64$ the error of the geometric
    integration still is smaller than the error of the
    time-parametrized integration at the highest tested resolution of
    $N_t=8192$ (overall $\approx 10^4$ times more
    accurate).\label{fig:burgers_convergence}}
\end{figure}
Figure~\ref{fig:burgers_convergence} shows the convergence of the
maximum of the Hamiltonian for the proposed scheme in comparison to
the time-parametrized variant. For both methods the Hamiltonian
converges to $0$ in second order. The reparametrized variant is a
factor $>10^4$ more accurate: For the smallest tested resolution of
$N_t=64$ the error of the geometric integration is still smaller
than the error of the time-parametrized integration at the highest
tested resolution of $N_t=8192$. The reason is simple:

\begin{figure}[tb]
  \begin{center}
    \includegraphics[width=0.7\textwidth]{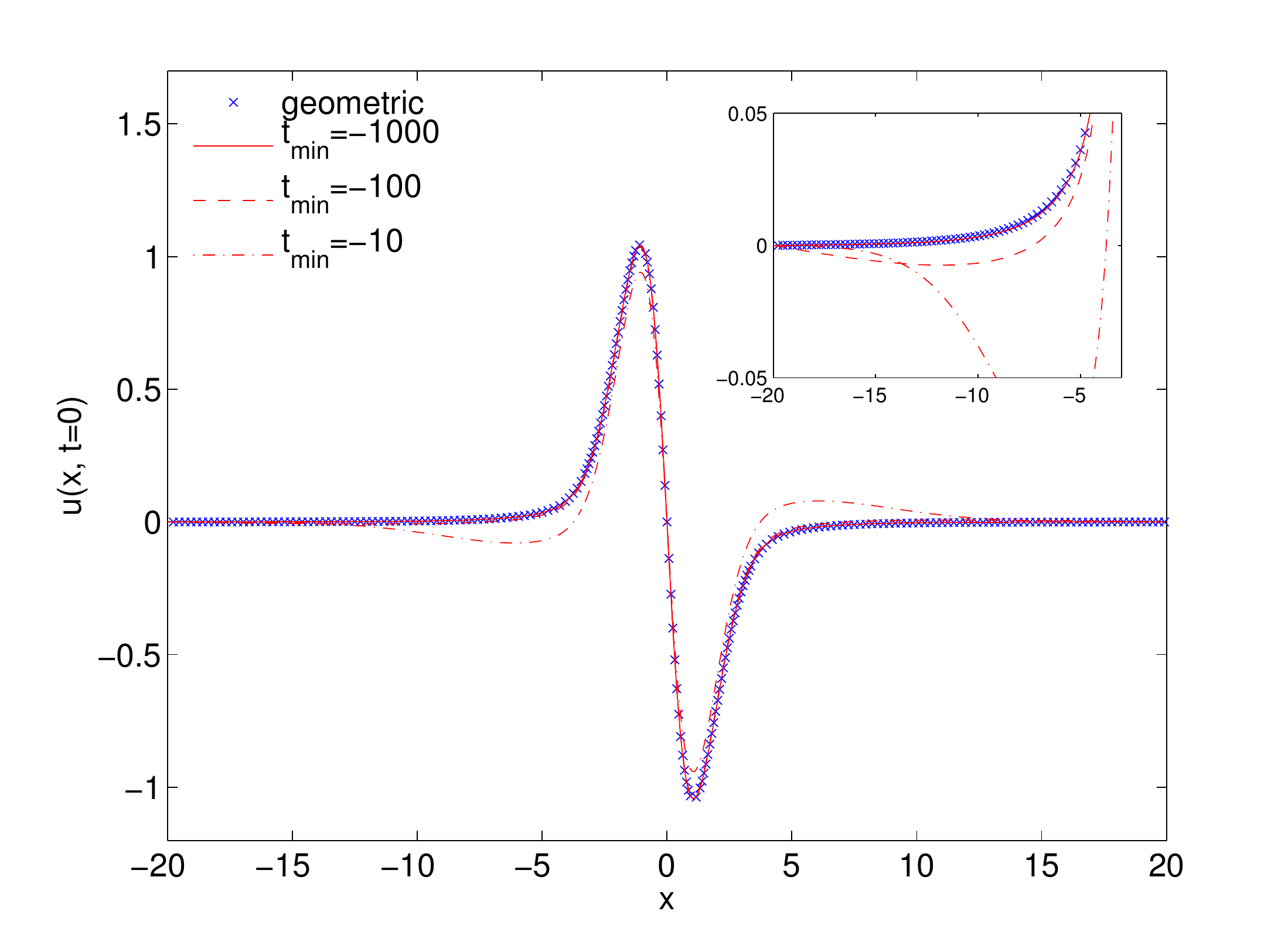}
  \end{center}
  \caption{Final configurations of the minimizer for different choices
    of the starting-time $t_{\text{min}}$ in comparison to the
    geometric variant. A choice of $t_{\text{min}}=-10$ and, to a much
    lesser extend, $t_{\text{min}}=-100$ produce secondary extrema
    (compare the zoom in the inset), which are absent for both
    $t_{\text{min}}=-1000$ and the geometric
    version.\label{fig:uends}}
\end{figure}
When using physical time as discretization, it is necessary
to prescribe a finite starting-time $t_{\text{min}}$ which is
restricted in its magnitude to ensure stability and accuracy of the
integration scheme for a given time resolution. Yet, when choosing
$t_{\text{min}}$ too close to $0$, the resulting finite-time minimizer
does not necessarily approximate the global minimizer. For the
geometric equations of motion this issue is absent, and the arbitrary
choice of $t_{\text{min}}$ is avoided in a natural
way. Figure~\ref{fig:uends} depicts the final configurations of the
minimizer for different choices of $t_{\text{min}}$, demonstrating the
necessity to choose a sufficiently small starting time: A choice of
$t_{\text{min}}=-10$ and, to a much lesser extend,
$t_{\text{min}}=-100$ produce secondary extrema, which are absent for
both $t_{\text{min}}=-1000$ and the geometric version.

\begin{figure}[tb]
  \begin{center}
    \includegraphics[width=0.48\textwidth]{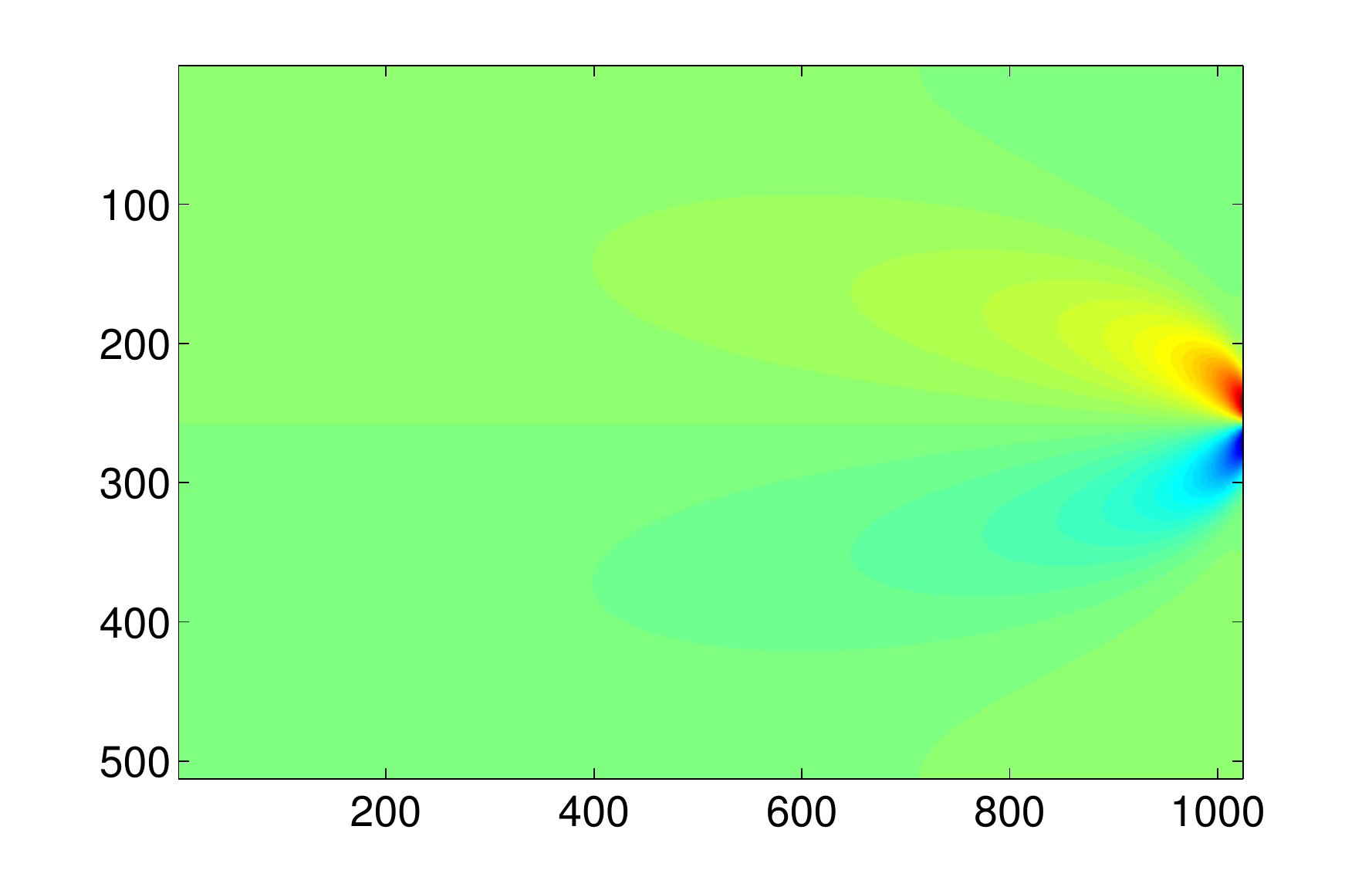}
    \includegraphics[width=0.48\textwidth]{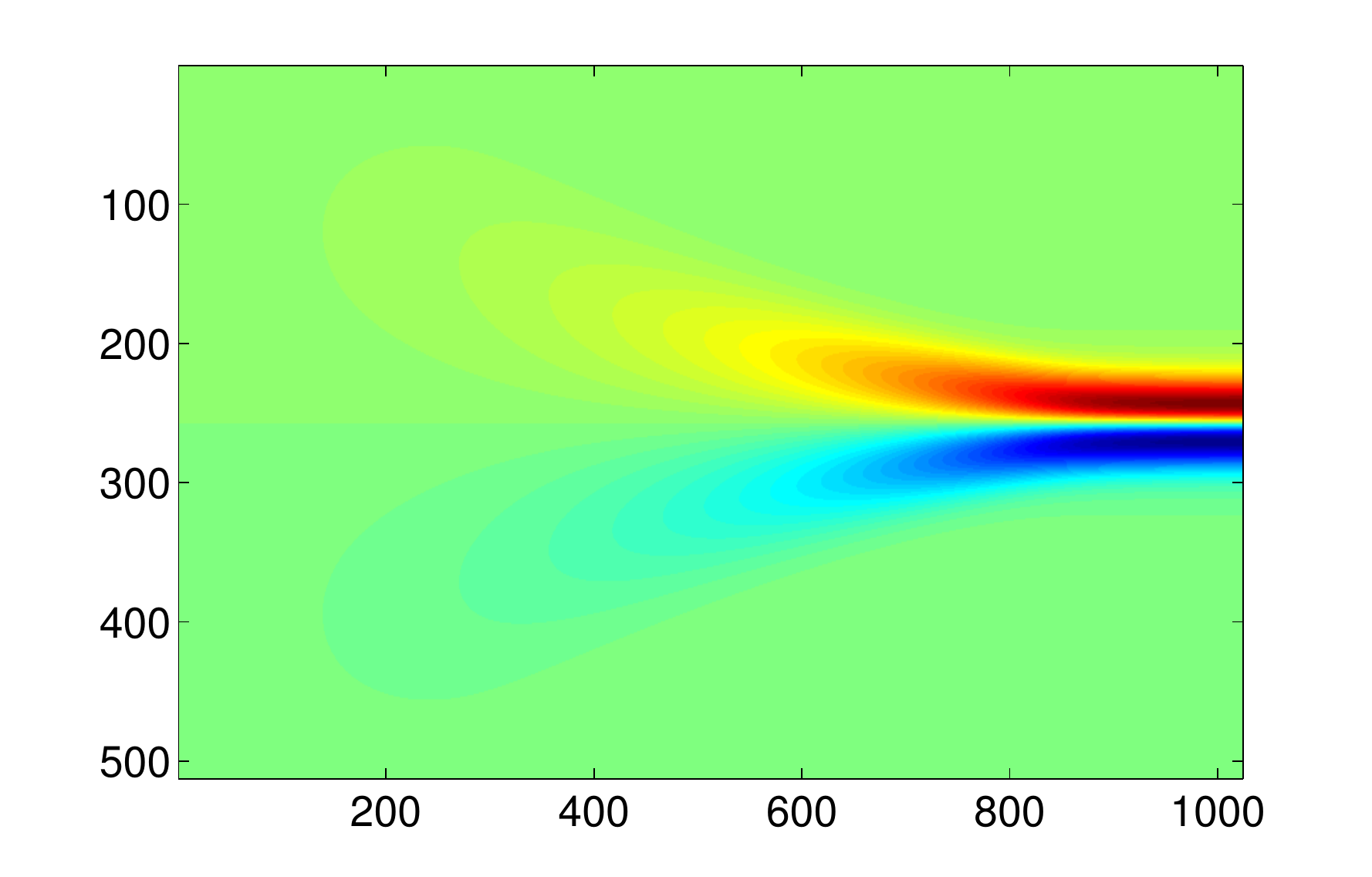}
  \end{center}
  \caption{Comparison of the minimizing field configuration for the
    time-parametrized equations of motion with $t_{\text{min}}=-100$
    (left) and the geometric equations of motion (right) in space and
    time. Most of the dynamics happen in a small interval near $t=0$,
    which is well resolved in the geometric case. \label{fig:spacetime}}
\end{figure}
A plot of the complete minimizer in space-time, as shown in
figure~\ref{fig:spacetime}, reveals the superior parametrization of
the geometric scheme. Even for a somewhat optimistic choice of
$t_{\text{min}}=-100$, most of the dynamics are squeezed into very few
grid-points for the time-parametrization (left), while being far
better resolved in the geometric case (right), which nevertheless
covers a more than 10 times larger interval of physical time in total.

\section{Conclusion} 
\label{sec:conclusion}

We presented a new method to compute minimizers of action functionals
using a geometric reparametrization of Hamilton's equations. The new
method is particularly well-suited to compute expectations related to
stochastically driven exits from attracting domains, though it may be
useful in other contexts as well. The new method was illustrated using
a simple Ornstein-Uhlenbeck system and then numerically implemented
and tested in the context of stochastic partial differential equations
with additive noise for the $\phi^4$-model and the stochastic Burgers
equation. In all cases, it was shown that the new parametrization has
computational advantages over the parametrization using the original
physical time. Applications to more complicated systems, in particular
higher-dimensional stochastic partial differential equations
(e.g. stochastically driven Navier-Stokes or MHD equations) are
therefore within reach and will be the topic of future research.

\section*{Acknowledgments}

The work of T.G. was partially supported through the grants
ISF-7101800401 and Minerva – Coop 7114170101. T.S. acknowledges
support through the following NSF grants: DMS-0807396, DMS-1108780,
and CNS-0855217. The work of E.V.-E. was supported in part by NSF
Grant DMS07-08140 and ONR Grant N00014- 11-1-0345.  Furthermore,
T.G. would like to thank Gregory Falkovich for interesting
discussions.

\end{document}